\documentclass[twoside,psfig,10pt]{article}
\usepackage{amsmath}

\title{ 
\Large
{\bf Mach-Zehnder interferometer based all optical flip-flop} 
}
\author{ 
{\bf Martin T. Hill, H. de Waardt, G. D. Khoe, H. J. S. Dorren} \; \;  
 \; \;
\\
{\small Department of Electrical Engineering, 
Eindhoven University of Technology,
P.O. Box 513,} \\
{\small 5600 MB Eindhoven, The Netherlands} }
\date{} 

\begin{document}
\maketitle

\begin{abstract} 
For the first time an all optical flip-flop is demonstrated based on two coupled Mach-Zehnder 
interferometers which contain semiconductor optical amplifiers in their arms. The flip-flop 
operation is discussed and it is demonstrated using commercially available fiber pigtailed 
devices. Being based on Mach-Zehnder interferometers, the flip-flop has potential for very 
high speed operation.
\end{abstract}

\vspace{3cm}
\begin{flushleft}
\begin{small}
\hspace{1cm}
E-mail: m.t.hill@ele.tue.nl
\end{small}
\end{flushleft}

\newpage
\section{Introduction}
Optical bistable devices and in particular all optical flip-flops can have many uses in optical
telecommunications and computing such as: threshold functions, 3R regeneration, 
de-multiplexing and rate conversion of telecommunication data [1].

All optical flip-flops based on two coupled devices provide many advantages such as 
controllable behaviour, separate set/reset inputs and identical set/reset operations, and large 
input wavelength range. Such a flip-flop using two coupled lasers was demonstrated and 
analyzed in [2]. In particular, the function of light output by a laser versus the light injected 
into a laser has the correct characteristics so that when two lasers are coupled the following 
occurs: The system has more than one steady state solution, and at least two of the steady state 
solutions are stable states of the system.

Integrated Mach-Zehnder interferometers (MZI) incorporating semiconductor optical 
amplifiers (SOA) in the interferometer arms have recently been developed as very high speed 
all optical switching devices [3],[4]. The function of light output by a MZI (with a constant 
bias light injected) versus the light injected into the MZI, can also have the correct 
characteristics for forming an all optical flip-flop. In this paper we experimentally demonstrate 
for the first time an all optical flip-flop based on two coupled MZI (with SOAs in their arms).

%
%
\section{Operating Theory}
The structure of the flip-flop is shown in Figure 1. It consists of two MZIs (MZI 1 and MZI 2) 
and a SOA connecting the two. The connecting SOA provides a controllable gain between the 
two MZIs. For the moment assume that the SOA gain is one and that it can be replaced with a 
direct optical connection between MZI 1 and MZI 2. Each MZI has a continuous wave (CW) 
bias light input \(S_{bias}\). 

Qualitatively the flip-flop functions as follows: With the flip-flop in state 1 light out of MZI 1 
flows into the SOA of MZI 2, changing gain and refractive index such that much less light 
exits from MZI 2 and flows back into MZI 1. State 2 is the reverse case where a large amount 
of light flows out of MZI 2 suppressing light flowing out of MZI 1 and back into MZI 2.

To switch the flip-flop between state 1 or 2 light can be injected into the MZI that dominates 
(that is the one injecting the most light into the other MZI), via the In 1 or In 2 ports (Figure 1). 
The injected light reduces the light exiting the dominate MZI, which allows the suppressed 
MZI to increase its light output and become the dominate MZI.

The flip-flop can be described quantitatively as follows. Each MZI can be modeled with: a rate 
equation for the carrier number of the SOA in its arms [5],[2], an equation relating carrier 
changes to refractive index and phase changes [6], and an equation to model the recombining 
of the signals in the arms at the MZI output coupler [4]. \(S_{out1}\) as a function of \(S_{in1}\) or 
equivalently \(S_{out2}\) can be found for the steady state from the MZI model for a particular set of 
operating conditions. One of these functions of \(S_{out1}\) versus \(S_{out2}\) is plotted in Figure 2. The 
SOA parameters used to construct the plot were from [6] (with the additional parameters of 
intrinsic losses $\alpha_{int} =27\:cm^{-1}$ and group velocity in the SOA  $\nu_g = 8\times 10^{9} 
cm\:s^{-1}$).

Also plotted in Figure 2 is \(S_{out2}\) as a function of \(S_{out1}\). The points where the two curves 
intersect represent steady state solutions for the system of two MZIs. The point labelled B1 
represents state 1 mentioned above. Point B2 represents state 2. Both B1 and B2 can be shown 
to be stable states of the system [2]. The point S represents a state where the same amounts of 
light flow from MZI 1 to MZI 2 and visa versa, however it is not a stable state of the system 
[2].

%
%
\section{Experiment}
To demonstrate the all optical flip-flop just described above the setup show in Figure 1 was 
realized using commercially available SOAs and fiber based couplers. The SOAs employed a 
strained bulk active region and were manufactured by JDS-Uniphase.

The central SOA was not necessary from a theoretical stand point. However it allowed the 
coupling between the MZIs to be varied as was necessary to obtain strong bistable operation. 

The amount of light being injected into each MZI 1 and MZI 2 was measured by a photo 
diodes PD 1 and PD 2, Figure 1. To toggle the flip-flop between states light pulses of power 
3.3 mW, wavelength 1547 nm, and duration 5 ms were regularly injected into the inputs In 1 
and In 2 (Figure 1). The CW bias light power was 1.25 mW, and had wavelength 1552 nm for 
MZI 1 and 1550 for MZ1 2. The MZI SOA currents were such that with only the CW bias light 
injected into them they had a gain of 10. 

The inputs pulses were injected every 50 ms into alternate MZIs. The changing of state of the 
flip-flop every 50 ms can be clearly seen in Figure 3, demonstrating proper flip-flop operation. 
Also the effects of the 5 ms input pulses can be seen. 

%
%
\section{Conclusions}
In this paper we have shown that it is possible to make an all optical flip-flop out of two 
Mach-Zehnder interferometers (with non-linear elements in their arms, here SOAs). Integrated 
versions of the flip-flop could operate at very high speeds [3], [4], as the MZIs themselves 
respond quickly and they are located close to each other. Furthermore the integrated MZIs will 
be stable. The use of MZIs makes the flip-flop presented here inherently faster than the 
flip-flop presented in [2] that is based on couplers lasers. The attributes of high speed and 
potentially wide input wavelength range would make the flip-flop suitable for all optical signal 
processing applications in high-speed telecommunications.

Other arrangements of the MZIs apart from that shown in Figure 1 are possible. For example it 
is possible to remove the SOA between the MZIs and share a common coupler between the 
MZIs. Additional couplers could be added in the MZIs for inputs or outputs. Finally the 
concept of optically bistable coupled MZIs could prove useful for increasing the sensitivity of 
MZI based optical sensors. 

\subsection*{Acknowledgments}
This research was supported by the Netherlands Organization for Scientific Research (N.W.O.) 
through the "NRC Photonics" grant.

\newpage
\subsection*{Figure Captions}
\vspace{10mm}
\noindent
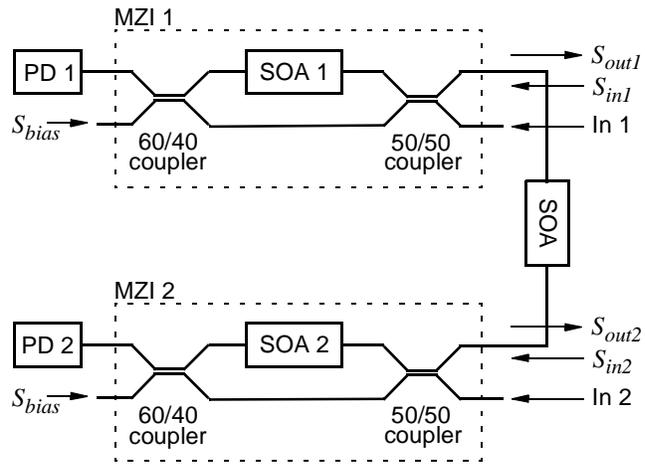
Figure 1: Structure of Mach-Zehnder Interferometer (MZI) based optical flip-flop. PD: photo diode, 
SOA: semiconductor optical amplifier

\vspace{10mm}
\noindent
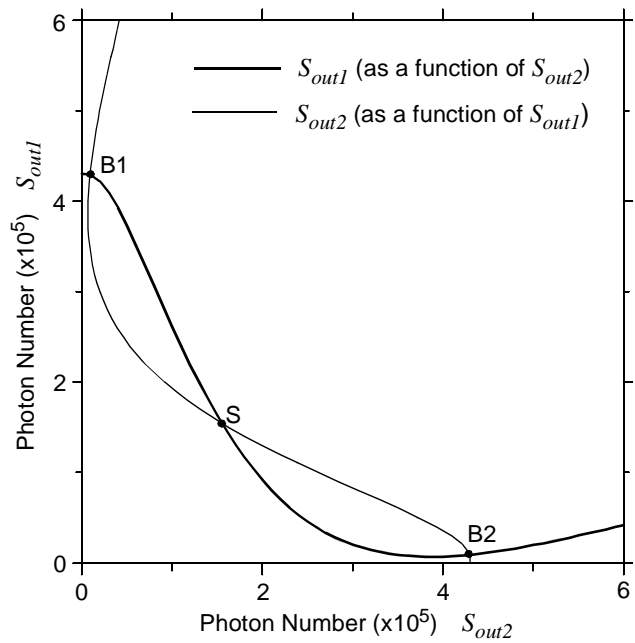
Figure 2: Steady state light output by a MZI as function of the light injected into it by the other MZI.

\vspace{10mm}
\noindent
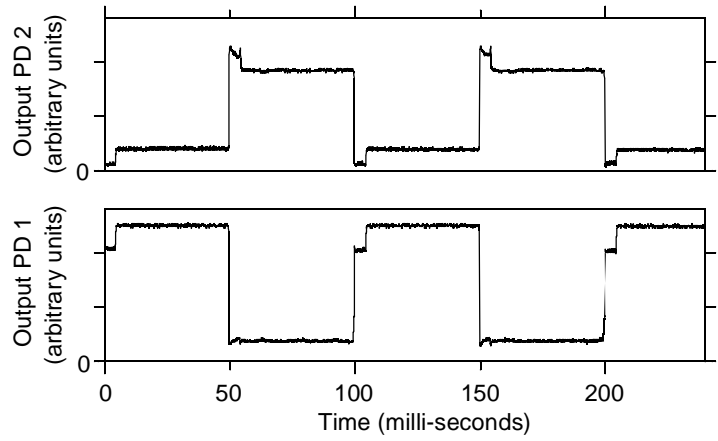
Figure 3: Oscilloscope traces of output of flip-flop showing switching between states every 50 
milli-seconds. Note that the effects of the 5 ms input pulses used to switch the flip-flop can also be seen in 
the traces.

\end{document}